\documentclass[12pt]{article}
\input epsf.sty
\newcommand{\be}{\begin{equation}}
\newcommand{\ee}{\end{equation}}
\newcommand{\ben}{\begin{eqnarray}\displaystyle}
\newcommand{\een}{\end{eqnarray}}
\newcommand{\refb}[1]{(\ref{#1})}

\def\sqr#1#2{{\vcenter{\vbox{\hrule height.#2pt
         \hbox{\vrule width.#2pt height#1pt \kern#1pt
            \vrule width.#2pt}
         \hrule height.#2pt}}}}

\begin{document}

{}~ \hfill\vbox{\hbox{hep-th/0307221} \hbox{PUPT-2091} }\break

\vskip 1cm

\begin{center}
\Large{\bf On the BCFT Description of Holes\\ in the {\bf $c=1$}
Matrix Model}

\vspace{10mm}

\normalsize{Davide Gaiotto, Nissan Itzhaki and Leonardo Rastelli}

\vspace{10mm}

\normalsize{\em Physics Department, Princeton University,
Princeton, NJ 08544}\end{center}\vspace{10mm}

\begin{abstract}

\medskip

We propose a
Boundary Conformal Field Theory description of
hole states in the $c=1$ matrix model.

\end{abstract}


\baselineskip=18pt

Recently McGreevy and Verlinde \cite{mv} showed that unstable
branes play a crucial role in the understanding of the duality
between $c=1$ matrix model and 2-d string theory (for
reviews see {\it e.g.} \cite{klebreview,
moorereview}). They suggested
that the matrix should be directly identified with the open string
tachyon field living on a large number of unstable branes of the
2-d string theory. This proposal was made very precise in
\cite{kms} (see also \cite{mtv,m,akk}) where it was shown that
Sen's rolling tachyon BCFT \cite{sen1} (tensored with the unstable
Liouville brane of \cite{zz}) is dual to a single matrix model
eigenvalue excited above the Fermi surface. These encouraging
results raise the question \cite{sen2}: is there a class of (time
dependent) BCFTs which describes the {\it holes} in the $c=1$
matrix model?

\medskip

If the duality between the $c=1$ matrix
model and 2-d string is correct the answer must be in the affirmative: 
holes are excitations with positive energy that should
have a dual description in the 2-d string theory.
Worries about the non-perturbative definition
of the theory (related to
eigenvalues tunnelling to the unfilled side of the potential)
should be even less relevant for the holes, which are
below the Fermi surface. We should expect to find a family of exact
BCFTs giving a classical description of the possible trajectories
of the hole \cite{sen2}.

\medskip

The aim of this short note is to answer this question. Our
starting point is Sen's exactly marginal boundary deformation that
describes the creation and decay of an unstable brane \cite{sen1},
\be \lambda \, \int \,
dt \, \label{iu} \cosh (t) \, . \ee
The boundary state associated with this deformation is
characterized by a time dependent function
\be\label{df}
f(t)=\frac{1}{1+e^t \sin(\pi\lambda
)}+\frac{1}{1+e^{-t} \sin(\pi\lambda )} -1 \, . \ee
The space-time interpretation is that
of incoming closed strings focusing
to form a D-brane at a time $t \sim -\tau$,
which then decays back into closed strings at $t \sim + \tau$,
with
\be \tau=-\log(\sin(\pi\lambda)) .\;\;\;\;\;\; \ee
The  energy of this solution is \be \label{energy}
 T_{00}=T_p \, \cos^2(\pi\lambda),\ee
where $T_p$ is the D-brane tension, which
in the context of  the matrix model is identified
with the cosmological constant
$\mu$. Note that for  $\lambda=\frac{1}{2}$  both the
energy and lifetime vanish. This special point, that was studied
in some detail in \cite{msy, llm, gir} in the critical string case,
corresponds in the matrix model to an excitation  right at
the Fermi surface. So this seems like the natural
starting point. Going above the Fermi surface ({\it i.e.},
creating an eigenvalue that goes up the potential and then rolls down)
corresponds to
\be\label{2} \lambda=\frac{1}{2}+\beta \, , \quad \beta \in {\bf R}  \, .\ee
The sign of $\beta$ does not affect any
physical quantity. Indeed the full boundary state is defined by
analytic functions of $(\lambda -\frac{1}{2})^2$. A simple way to
see this is to Wick rotate back to the Euclidean BCFT
\cite{callan, pol} with boundary deformation $\lambda \cos(X)$.
At $\lambda = \frac{1}{2}$, this BCFT describes an array of 
D-branes and the operator $\cos(X)$ is associated with open strings
streched between neighboring branes; hence an {\it even} number of
insertions of $\cos(X)$ is needed to obtain non-vanishing disk
amplitudes.

\medskip

Note that the time delay, $2 \tau$, between the incoming and
outgoing closed string radiation is positive. In the context of
the matrix model the value of $\tau$ is in  precise agreement
with the time delay of a classical trajectory of energy
\refb{energy} with respect to the trajectory at the Fermi level
\cite{kms}.

\medskip

If the deformation (\ref{2}) corresponds to going above the Fermi
surface, then by continuity one might suspect that the deformation
corresponding to going below the Fermi surface is
\be\label{def} \left(\frac{1}{2} + i \alpha \right) \, 
\int \, dt \, \cosh (t)\, ,
\,\;\;\;\; \;\;\;\;\;\alpha \in {\bf R} \,. \ee
This may at first look like an implausible
attempt since this is a complex deformation of the worldsheet
action. However, as remarked above, the boundary state is a
function of $(\lambda -\frac{1}{2})^2$, and is thus real.

\medskip

For example, $f(t)$, that
fixes the closed string "tachyon" in the 2-d case and many of the
closed strings fields in the critical string case, is
real,
\be  f(t)=\frac{1}{1+e^t \cosh(\pi \alpha )}+\frac{1}{1+e^{-t}
\cosh(\pi \alpha)} -1 \, . \ee
The time delay and energy are
\be\label{34} \tau=-\log(\cosh(\pi\alpha)) \, ,
\quad T_{00}=-\mu \sinh^2(\pi \alpha) \, .
\ee
Note that both are negative. We would like to argue that this is
precisely as it should be. We claim that the boundary state with
deformation \refb{def} corresponds to the {\it classical}
trajectory of an {\it eigenvalue}  below the Fermi surface. This
explains the negative value of the energy. The negative value of
$\tau$ is also in agreement with the matrix model since now there
is a time delay between a classical trajectory at the Fermi
surface and the classical trajectory {\it below} the Fermi surface. Indeed
one can easily generalize the results of \cite{kms} to find a
precise agreement between the matrix model time delay and
\refb{34}.

\medskip

In the quantum description, we must associate with this negative
energy solution a {\it destructor} operator.  The hole state will
then have positive energy and it will couple to the closed string
fields with {\it minus} the couplings computed from the
deformation \refb{def}.  We are now in the position to state
precisely our conjecture.  The boundary state associated with a
hole in the $c=1$ matrix model is obtained by tensoring the
$(1,1)$ Liouville brane of \cite{zz} with {\it minus} the
``boundary state'' associated with the boundary deformation
\refb{def}.

\medskip

As an immediate test of this proposal, let
us compute the  outgoing closed string tachyon
radiation associated with our candidate hole state.
If we {\it did not} include an extra minus sign
in front of the boundary state,
the outgoing radiation associated
with the deformation \refb{def} would take the
form \cite{kms}
\be\label{89} |\psi (\tau) \rangle \sim \exp\left( \int_{0}^\infty \frac{dp}{\sqrt{E_p}}\,
a^{\dag}_p A(p)\right) | 0 \rangle ,\;\;\;\; \mbox{where}\;\;\;\;\;A(p) \sim i
\exp(i E_p \tau). \ee
This works just fine  to describe the creation of an eigenvalue
\cite{kms} since bosonization implies that the
fermion creation operator is
\be \psi^\dagger = :\exp(i  \phi ): \,.\ee
Now for
$\lambda = \frac{1}{2} + i \alpha$ the time delay $\tau $ is
negative but this does not change the sign in the exponent (as can be seen
from (\ref{89})), so the BCFT \refb{def} is  still describing the
creation of an eigenvalue rather then
the creation of a hole. The negative time delay $\tau$ tells
us that this eigenvalue is below the Fermi surface.
Multiplying the boundary state by an overall
minus sign flips the sign of $A(p)$, and gives
precisely the bosonization formula for a hole, $\psi = :\exp(-i \phi): $.

\medskip

 An immediate extension of this result
is to consider the $(\pm \frac{1}{2} + i \alpha) \cosh$
deformation in superstring theory. Using the formulae
in \cite{sen1,msy} it is easy to check that the boundary state
is again real, including the RR couplings. The interpretation
is that {\it minus} this boundary state describes
holes in the $\hat{c} =1$ matrix model \cite{tt,6}. 

\medskip
While the motivation to consider the boundary deformation \refb{def} 
came from the
duality between the matrix model and 2-d string theory, 
it is very tempting to consider it in the context of 
critical string theory. Since as $\alpha \to \infty$ the energy diverges, 
by taking large
$\alpha$ while keeping $g_s$ small but finite the
system will collapse to form a Schwarzschild black hole. So this
deformation might teach us something new about the relation
between strings,  branes and black holes. It is interesting to note
that, much like in the black hole case, the lifetime associated
with our deformation grows with the energy. However, it
grows only logarithmically with the energy. Of course this is just the
classical open string result, which is valid at $g_s=0$ and hence it
is not aware of any black hole physics. It should be interesting
to see whether loop effects will increase that lifetime in
such a way that an agreement will be found at the transition point
of \cite{hp}.

\medskip
We would like to end with an amusing observation about the supersymmetric case.
In this case $f(t)$ takes a slightly different form \cite{sen1},
\be\label{dfs} f(t)=\frac{1}{1+e^t \sin^2(\pi\lambda
)}+\frac{1}{1+e^{-t} \sin^2(\pi\lambda )} -1 \, . \ee
Notice that now besides $\lambda = \pm \frac{1}{2} + i \alpha$
there are two other complex boundary deformations
that yield a real $f(t)$,
\be \label{ialpha} i\alpha \cosh(t)\;\;\;\;\mbox{and}\;\;\;\;i\alpha
\sinh(t)\,, \;\; \quad \alpha \in {\bf R} \,.\ee
These deformation have an interesting physical meaning. Let us
focus for example on the first. In this case we get
\be  f(t)=\frac{1}{1-e^t \sinh^2(\pi \alpha )}+\frac{1}{1-e^{-t}
\sinh^2(\pi \alpha)} -1. \ee
Unlike the deformation \refb{def}, that gave a smooth boundary state,
now we find that $f(t)$ 
is singular at $t = \pm t_0$, with
$t_0= \log(\sinh^2(\pi\alpha))$. Consider $|\alpha| \ll 1$.
For $|t|>|t_0|$ we see that $f(t)$ goes to
zero while for $|t|<t_0$ it goes to one. Since $f=0$ corresponds to
the vacuum and $f=1$ corresponds to the creation of the unstable
brane, this deformation describes exactly the
configuration considered in \cite{dgi}. Namely we have an unstable
D0-brane (in, say, 10-d IIB) whose worldline starts on one-half
D-instanton, located at the singular point $t= - t_0$,
and ends on an {\it anti} one-half D-instanton
located at $t = + t_0$. 
By considering the
RR-fields one can verify that the second deformation in \refb{ialpha} 
corresponds
to an unstable D0-brane that starts on one-half D-instanton and
ends on another one-half D-instanton. It should be interesting to
see if this realization could shed new light on the 
somewhat mysterious physics of merons in gauge theories \cite{cdg}.


\bigskip

\bigskip

\noindent {\bf Note Added:} The main observation of the present
paper is also made in \cite{6} that has just appeared on the archive.

\bigskip

\bigskip

\noindent {\bf Acknowledgements}

We thank A. Sen for useful discussions.
This material is based upon work
supported by the National Science Foundation under Grant No. PHY
9802484 and PHY-0243680. Any opinions, findings, and conclusions
or recommendations expressed in this material are those of the
author and do not necessarily reflect the views of the National
Science Foundation.


\end{document}